\documentclass[aps,prb,groupedaddress,showpacs,twocolumn,superscriptaddress]{revtex4}
\usepackage{graphicx}
\usepackage{amsmath}
\usepackage{amssymb}
\usepackage{bbm}
\usepackage{xspace}
\usepackage{color}
\usepackage{footnote}
\usepackage{comment}
\usepackage{placeins}
\usepackage{hyperref}
\usepackage{tcolorbox}
\usepackage{dsfont}

\newcommand{\eq}[1]{Eq.\thinspace{}(\ref{#1})}
\newcommand{\Eq}[1]{Eq.\thinspace{}(\ref{#1})}

\newcommand{\etal}[0]{\textit{et al.}}
\newcommand{\tcite}[1]{Ref.~\onlinecite{#1}}

\newcommand{\iim}{\Im{}\,}
\newcommand{\li}{\mathcal L}
\newcommand{\ga}[1]{\Gamma^{(#1)}}

\newcommand{\vv}[1]{\boldsymbol{#1}}

\newcommand{\nag}{{\phantom{\dagger}}}

\newcommand{\aux}{\mathrm{aux}}
\newcommand{\ph}{\mathrm{ph}}

\def\bra#1{\mathinner{\langle{#1}|}}
\def\ket#1{\mathinner{|{#1}\rangle}}
\def\braket#1{\mathinner{\langle{#1}\rangle}}

\def\nr2#1{\mathinner{\lvert\lvert{#1}\rvert\rvert^2}}

\definecolor{orange}{RGB}{252,77,6}
\definecolor{brown}{RGB}{200,127,50}
\definecolor{green1}{RGB}{00,100,00}
\definecolor{green2}{RGB}{00,150,00}
\definecolor{green3}{RGB}{00,200,00}

\usepackage{ulem}

\begin{document}

\title{Auxiliary master equation approach within stochastic wave functions: Application to the Interacting Resonant Level Model}

\author{Max E. Sorantin}
\email[]{sorantin@tugraz.at}
\affiliation{Institute of Theoretical and Computational Physics, Graz University of Technology, 8010 Graz, Austria}
\author{Delia M. Fugger}
\affiliation{Institute of Theoretical and Computational Physics, Graz University of Technology, 8010 Graz, Austria}
\author{Antonius Dorda}
\affiliation{Institute of Theoretical and Computational Physics, Graz University of Technology, 8010 Graz, Austria}
\author{Wolfgang von der Linden}
\affiliation{Institute of Theoretical and Computational Physics, Graz University of Technology, 8010 Graz, Austria}
\author{Enrico Arrigoni}
\email[]{arrigoni@tugraz.at}
\affiliation{Institute of Theoretical and Computational Physics, Graz University of Technology, 8010 Graz, Austria}

\date{\today}

\begin{abstract}
We present further developments of the auxiliary master equation approach (AMEA), a numerical method to simulate many-body quantum systems in as well as out of equilibrium, and apply it to the Interacting Resonant Level Model (IRLM) to benchmark the new developments. In particular, our results are obtained by employing the stochastic wave functions (SWF) method to solve the auxiliary open quantum system arising within AMEA.
This development allows to reach extremely low wall-times for the calculation of correlation functions with respect to previous implementations of AMEA.
An additional significant improvement is obtained by extrapolating a series of results obtained by increasing the number of auxiliary bath sites, $N_B$, used within the auxiliary open quantum system formally to the limit of $ N_B \rightarrow \infty$.
Results for the current-voltage characteristics and for equilibrium correlation functions
are compared with the one obtained by exact and matrix-product states based approaches.
\end{abstract}


\maketitle

\section{Introduction}\label{sec:introduction}
Quantum impurity models have a long history in many-body quantum mechanics. Some prominent examples are, the Single Impurity Anderson Model\cite{ande.61} (SIAM), the (Anderson-) Holstein Model\cite{ho.59},the Kondo Model\cite{kond.64} and the Interacting Resonant Level Model\cite{wi.fi.78} (IRLM). They feature interesting, unconventional physics such as the Kondo effect\cite{hews} or negative differential conductance\cite{bo.sa.08} and allow for experimental realizations in terms of quantum dots\cite{go.go.98}. 
Besides this, the solution of quantum impurity problems alone constitutes already a crucial task in Dynamical Mean Field Theory\cite{me.vo.89}.
\\
Over the last decade, there has been increasing interest in quantum impurities out of equilibrium and the development of numerical methods which are able to accurately simulate such systems poses a great challenge for contemporary condensed matter theory. Existing methods~\cite{ec.ha.09} include, iterated perturbation
theory~\cite{sc.mo.02u}, numerical renormalization group~\cite{jo.fr.08},
real time  quantum monte carlo (QMC)~\cite{ec.ko.09,ec.ko.10},
noncrossing approximation and beyond~\cite{okam.08,ar.ko.12}, 
or imaginary-time QMC supplemented by a double analytical
continuation~\cite{han.07,ha.he.07,di.we.10,ar.we.13}, scattering-states approaches~\cite{me.an.06,ande.08}, perturbative and renormalization group(RG) methods~\cite{me.wi.92,sc.sc.94,ro.pa.05,scho.09}, time-dependent density-matrix RG (DMRG) and related tensor-network approaches~\cite{wh.fe.04,da.ko.04,pr.zn.09}, numerical RG~\cite{an.sc.05}, flow equation~\cite{kehr.05},  functional RG~\cite{ge.pr.07,ja.me.07}, dual fermions~\cite{ju.li.12,ch.co.18u}.
A method developed over the last years is the so-called auxiliary master equation approach\cite{ar.kn.13,do.nu.14,do.so.17} (AMEA). The advantage of this approach is that, 
in contrast to approaches which simulate a closed Hamiltonian system, it allows to directly address the steady state. Also time-dependent correlation functions can be readily evaluated starting from the steady state or any arbitrary initial condition. AMEA was successfully used as impurity solver within steady state non-equilibrium DMFT\cite{ti.do.15,do.ti.16,ti.do.17,ti.so.18} as well as to calculate highly accurate spectral functions of the SIAM under the influence of a bias voltage\cite{do.ga.15,fu.do.18}.\\
AMEA is based upon mapping the physical system to an auxiliary open quantum system of Lindblad form. The dynamics of the resulting auxiliary system is described by the density matrix and is solved by numerical means. In previous works the Lindblad system was solved by using the so called super-fermion (SF) representation\cite{dz.ko.11}, which formulates the super operator problem in terms of a standard operator problem with twice as many sites. The operator problem was than solved by standard numerical many-body techniques such as Krylov-space methods\cite{kn.ar.11.ec,do.nu.14} (ED) or time evolution with Matrix Product States\cite{do.ga.15} (MPS).\\
In this work, we implement an alternative solution strategy which does not rely on the SF representation, namely Stochastic Wave functions\cite{da.ca.92,br.ka.97,br.ka.98} (SWF). The new method is statistical in nature and most notably highly parallelizable. This makes it a very promising candidate to exploit the multi-core architecture of (future-) cluster facilities. In addition, we introduce the notion of finite-size scaling within AMEA and report on progress regarding the optimization problem arising when mapping to the auxiliary system.\\
To test and benchmark the new developments, we apply AMEA to the IRLM where we can compare to existing literature. This work is structured as follows.\\
We begin by describing the technical aspects in Sec.~\ref{sec:AMEA}-\ref{sec:ExtrapolationofObservables} and present the results for the IRLM in Sec.~\ref{sec:ApplicationtotheIRLM}. In more detail, in Sec.~\ref{sec:AMEA} we outline AMEA for spinless one dimensional systems. Sec.~\ref{sec:SWF}, and the corresponding appendix Sec.~\ref{app:SWF} , is devoted to the description of the SWF algorithm, the finite-size scaling is introduced in Sec.~\ref{sec:ExtrapolationofObservables}. In Sec.~\ref{sec:ApplicationtotheIRLM} we apply AMEA within SWF to the IRLM and test the finite-size scaling scheme as well as the capability to compute correlation functions against the literature. Finally, we present our conclusion together with a summary and outlook in Sec.~\ref{sec:Conclusion}

\section{Auxiliary Master Equation Approach}\label{sec:AMEA}
We briefly review  AMEA to deal with fermionic impurity problems. We consider a generic interacting region - the impurity - of size $N_{\text{imp}}$ connected to a left and right bath of non-interacting fermions.
Accordingly, we write the Hamiltonian as
\begin{equation}
 H=H_{\text{imp}} + H_{\text{Baths}} + H_{\text{Hyb}}\, .
 \label{eq:splitHamiltonian}
\end{equation}
Here, $H_{\text{imp}}$ describes the interacting region, $H_{\text{Baths}}=\sum_{\alpha=L/R} H_{\text{B}_\alpha}$ corresponds to the leftover reservoirs and $H_{\text{Hyb}}$ contains the hopping terms connecting the baths to the impurity. In the following we will assume that an individual bath is connected only to a single site of the impurity.\\
The idea of AMEA is now to model the physical situation by an auxiliary open quantum system described by the Lindblad equation. It consists of the impurity and additional bath sites to approximate the action of the leftover Hamiltonian on the interacting region.\\
In more detail, the Lindblad super-operator (Liouvillian) defining the 
dynamics of the open quantum system of size $L=N_{\text{imp}} + 2N_B$ reads\footnote{for simplicity we neglect spin}
\begin{align}
 \mathcal{L}\rho &= -i[H_{\text{imp}},\rho] + \mathcal{L}_{D}\rho\\
 \mathcal{L}_{D}\rho &= \sum_{\alpha=L/R}\mathcal{L}_{\alpha}\rho\, ,
 \label{eq:Lindbladian}
\end{align}
where $\rho$ is the density matrix of the Lindblad system. The Liouvillian of the dissipative bath sites is given by
\begin{align}
\notag \mathcal{L}_{\alpha}\rho &= -i\sum_{ij}E_{ij}^{(\alpha)}\left[ c_i^{\dagger}c_j,\rho\right] \\
\notag +& 2\sum_{ij}\Gamma_{ij}^{(\alpha),(1)}\left(c_j\rho c_i^{\dagger} - \frac{1}{2}\left\{\rho,c_i^{\dagger}c_j\right\}\right)\\
 +& 2\sum_{ij}\Gamma_{ij}^{(\alpha),(2)}\left(c_i^{\dagger}\rho c_j - \frac{1}{2}\left\{\rho,c_jc_i^{\dagger}\right\}\right)\, ,
 \label{eq:BathLindbladian}
\end{align}
where $\alpha$ denotes the left/right reservoir\footnote{Here, it is worth noting that the matrices $E^(\alpha),\Gamma^{(\alpha),1},\Gamma^{(\alpha),2}$ are only non-zero in the part of the system which describes the corresponding bath.} and $c_i^{(\dagger)}$ are the creation (annihilation) operators of a fermion on site $i$ of the open quantum system. The time-evolution of the system is described by the Lindblad equation,
\begin{equation}
 \frac{d}{dt}\rho(t) = \mathcal{L}\rho(t)\, .
\end{equation}
For the steady state of the original system, Eq.\ref{eq:Hamiltonian}, the Dyson equation for the interacting region in the formulation of Keldysh Green's functions reads,
\begin{equation}
 \mathbf{G}_{\text{imp}}^{-1}(\omega) = \mathbf{g}_{0,\text{imp}}^{-1}(\omega) - \mathbf{\Delta}_{\text{ph}}(\omega) - \mathbf{\Sigma}(\omega)\, ,
 \label{eq:impuritDyson}
\end{equation}
where all objects have a matrix structure in the physical sites and Keldysh-space. In Eq.~\ref{eq:impuritDyson} $g_{0,\text{imp}}$ is the Green's Function (GF) of the interacting region when isolated from the baths and without interaction,
$\Sigma$ 
is the unknown selfenergy, holding all information about the interaction and $\Delta$ is the so-called hybridization describing the effect of $H_{\text{Baths}} + H_{\text{Hyb}}$ on the impurity. For the case considered here, the hybridization has the spacial structure $\text{diag}(\Delta_{\text{ph}}^{(L)},0,...,0,\Delta_{\text{ph}}^{(R)})$.\\
The mapping from the physical to the auxiliary system is performed by fitting the parameters $E^{\alpha},\Gamma^{\alpha,(1,2)}$ in Eq.~\ref{eq:BathLindbladian} such that the hybridization in the auxiliary system approximates the physical hybridization as close as possible, $\Delta_{\text{aux}}^{(\alpha)}\approx \Delta_{\text{ph}}^{(\alpha)}$, and this is the only approximation made within AMEA. The accuracy of the mapping can than be systematically improved by increasing the number of auxiliary bath sites $N_B$ and it becomes formally exact in the limit of $N_B \rightarrow \infty$. Once the mapping is performed, one can solve the auxiliary system by some appropriate numerical method and evaluate observables belonging to the impurity. Their accuracy in describing the corresponding exact quantities will be directly related to the difference between
$\Delta_{\text{aux}}^{(\alpha)}$ and $\Delta_{\text{ph}}^{(\alpha)}$.

\subsection{Mapping to the auxiliary System}\label{subsec:MappingtotheauxiliarySystem}
Here, we briefly want to  summarize the mapping procedure and mention key points that we need for the present work. For a thorough discussion of the mapping and technical details we refer to our previous work\cite{do.so.17}. The mapping is performed for each individual bath $\alpha$
 by minimizing a suitable cost function
\begin{eqnarray}
\chi^2(\vv{x}_{\alpha})\equiv \chi_{\alpha}^2 &=&\int\left| \Delta_\ph^{(\alpha)} - \Delta_\aux^{(\alpha)} \right| d\omega \, , \nonumber \\
\left| \Delta_\ph^{(\alpha)} - \Delta_\aux^{(\alpha)} \right| &=& \hspace{-0.9em} \sum\limits_{\xi\in\{\text{Ret,Kel\}}}\hspace{-0.9em} \left[ \iim\Delta^{(\alpha),\xi}_\ph(\omega) - \iim\Delta^{(\alpha),\xi}_\aux(\omega;\vv{x}_{\alpha}) \right]^2  \, .
\label{eq:costfunc}
\end{eqnarray}
Here we have introduced a parameter vector $\vv{x}_{\alpha}$ that 
parametrizes the 
 matrices $E^{\alpha},\Gamma^{\alpha,(1,2)}$ in \eqref{eq:BathLindbladian}, from which one evaluates the auxiliary hybridization $\Delta_\aux^{(\alpha)}$. It is important to note that the precise form of the cost function is very flexible and may be chosen differently for different physical situations. One important property of the mapping is that the cost function decreases exponentially with the number of fit parameters, 
$ -\log\chi_{\alpha} \propto \text{dim}(\vv{x_{\alpha}})$, which typically leads to a rapid increase of accuracy when the number of bath sites $N_B$ is increased.

In previous works, Eq.~\ref{eq:costfunc} was minimized via a parallel tempering (PT) algorithm which is appropriate to  find the global minimum. However, it should be noted that within AMEA it is not strictly necessary to find the global optimum\footnote{although desirable as it gives the best approximation for a given system size.}. In general, the fit struggles to resolve sharp features such as band-edges in the retarded component or the fermi-jumps in the Keldysh component at zero temperature. Therefore, $T=0$ can not be reached exactly in practice and the auxiliary system always has some non-zero effective temperature.

\subsubsection{Developments of the fit}\label{subsubsec:fit}
With increasing dimensionality of the fitting problem, the PT algorithm gets
computationally prohibitive and it is not able to find even good local minima anymore for\footnote{This corresponds to the case $N_B>6$ when allowing for the most general Lindblad couplings.} $\text{dim}(\vv{x})=2N_B(N_B+1)\gtrsim 80$. 
Good minima should be such that they display an exponential decrease in the cost function when the number of bath sites is increased. To obtain good enough minima for $N_B=7,8$, we use the fact, which we observed empirically, that the $\Gamma$ matrices of obtained minima typically have very low rank. Utilizing a variable rank parametrization in terms of a corresponding matrix $H$
\begin{equation}
 H = (\vec h_1,..,\vec h_{\text{rank}_H}), \hspace{10pt}\Gamma = HH^{\dagger}\, ,
 \label{eq:vectorizedparametrization}
\end{equation}
where $\vec h_i$ denote column vectors of length $L$.
 Note that the maximal useful rank typically increases with the system size\footnote{For example, we observed that for a system with $N_B=9$ bath sites, increasing $\text{rank}_H>4$ was not fruitful in terms of the cost-function.} 
With this procedure, we have reduced the dimensionality of the parameter vector to $\text{dim}(\vv{x})=2N_B(\text{rank}_H + 1)$ extending the applicability of the PT algorithm to about $N_B=8$. To achieve an exponential decrease in the cost function for even more bath sites we have adopded an optimization algorithm
which makes use of the gradient of the cost function, which can be evaluated directly. This information is not used in the PT algorithm.
Suitable gradient-based approaches can be found in the area of machine learning, which provides algorithms tailored to find local minima in very high-dimensional problems utilizing variants of steepest descent. Here, we employ the ADAM~\cite{ADAM} optimizer as implemented in the python library tensorflow\cite{tensorflow}.\\
Steepest descent approaches are obviously very sensitive to the  starting point. In our case,  it has proven to be very effective to first find the solution for  a small auxiliary system (small $N_B$) and consequently add bath sites until the required $N_B$ is reached. For a fixed $N_B$ we start with the result of the previous system size and increase the rank stepwise until no significant decrease in the cost-function is observed. In addition to beeing applicable for larger $N_B$, the ADAM routine is faster than PT for a given $N_B$.\footnote{This is because the PT algorithm tries to explore the total phase space, wheres as ADAM only follows a certain path.}.

\section{Solution of the Linblad system with Stochastic Wave functions}\label{sec:SWF}
The auxiliary open system is still correlated  but due to its finite size
can be addressed by numerical techniques. 
One route is to make use of the so-called Super-Fermion (SF) representation \cite{dz.ko.11}, which maps a super-operator problem to a standard, albeit non-hermitian,  operator problem.
The drawback of this approach is that the resulting SF problem is formulated on twice as many effective sites leading to a rapid increase in the 
numerical complexity.
In  previous works employing AMEA we have successfully used the SF representation together with established many-body techniques such as Krylov-space methods\cite{ar.kn.13,do.nu.14} or MPS\cite{do.ga.15} to solve for steady state properties.
A completely different route is to use Stochastic Wave functions\cite{da.ca.92,br.ka.97,br.ka.98} (SWF), also referred to as ``quantum jumps'', to solve the auxiliary many-body problem.
The method is based on the stochastic nature of the Lindblad problem and is formulated in terms of wave functions instead of a density matrix and thus circumvents the need to square the  Hilbert space. In the following, we will only give a brief introduction to the SWF method and focus more on a practical prescription to simulate the many-body Lindblad system arising within AMEA. For more details, mathematical definitions and background we refer to the literature\cite{da.ca.92,br.ka.97,br.ka.98}.\\
The density operator $\rho(t)$ can be mapped onto a probability distribution $P[\psi(\lambda),t]$ for the quantum mechanical (many-body) wave function\footnote{To be consistent with quantum mechanics, $P[\psi,t]$ must not depend on the phase of the wavefunction and it is only non-vanishing for normalized states.}
\begin{equation}
 |\psi\rangle = \sum_{\lambda} \psi(\lambda)|\lambda\rangle,
 \label{eq:psi_MB_expansion}
\end{equation}
where $\lambda$ indexes a complete set of (many-body) basis states\footnote{More generally, $\lambda$ indexes a complete set of quantum numbers}. With the Hilbert space volume element,
\begin{equation}
  D\psi D\psi^* \equiv \prod_{\lambda}\frac{i}{2}d\psi(\lambda)d\psi^*(\lambda)\, ,
  \label{eq:HspaceVelement}
\end{equation}
defining the needed probability measure\footnote{$P[\psi(\lambda),t]D\psi D\psi^*$ can then be interpreted as the probability to find the system within the volume element $D\psi D\psi^*$ around the state $\psi(\lambda)$ at time $t$.}, the expectation value of an observable can then be formally expressed as
\begin{equation}
 \langle A(t)\rangle = \int D\psi D\psi^*\langle\psi|A|\psi\rangle P[\psi(\lambda),t] \, .
 \label{eq:QJObservable}
\end{equation}

In short, instead of dealing with an evolution equation for the density matrix, one formulates a stochastic process on the Hilbert space.\\
For the specific case of a Lindblad system, the process is simulated according to a stochastic differential equation\footnote{The mapping to a stochastic differential equation is possible also in a more general context, but the exact form of the latter is only known in special cases.} leading to the algorithm presented in Fig.~\ref{fig: SWFalgorithm}.

\begin{figure}
\begin{tcolorbox}[title=Stochastic wave function (SWF) algorithm, halign title=flush center, halign lower=center, fonttitle=\bfseries, lower separated=false, colframe=blue!70!white, colback=blue!5!white, left=0mm,  leftlower=4mm, sharp corners]
  \begin{enumerate}
    \item Start with a normalized state $\ket{\psi(t_0)}$ and draw a random number $r_\mathrm{j}\in(0,1)$.
    \item Time evolve the state vector with the effective Hamiltonian Eq.~\ref{heff}: $\ket{\psi(t)} = e^{-i H_\mathrm{eff}(t-t_0)}\ket{\psi(t_0)}$ up to a time $t_j$ such that $\nr2{\psi(t_j)} = r_j$.
    \item Perform a quantum jump:
    \begin{itemize}
      \item Compute the  weights for all possible jumps, $w_{\beta k} \propto \nr2{ L_{k}^{(\beta)}\psi(t_j) }$. 
      \item Select one jump process $(\beta'k')$ at random according to the weights.
      \item Change $\ket{\psi(t)} = L_{k'}^{(\beta')}\ket{\psi(t_j)}$ and normalize it. 
    \end{itemize}
    \item Set $t_0=t_j$ and iterate $1\to4$.
  \end{enumerate}
  \tcblower
  \tcbsubtitle[after skip=\baselineskip]{\centering Single-time observables}
  Measure at desired times and average over a sufficient number of realizations $\psi_i(t)$.
  \begin{equation*}
    \braket{A(t)} = \frac{1}{n} \sum_i^n \frac{\braket{\psi_i(t)\rvert A\lvert\psi_i(t)}}{\nr2{\psi_i(t)}}
  \end{equation*}
  
\end{tcolorbox}
\caption{The stochastic wave function algorithm for the time evolution.}
\label{fig: SWFalgorithm}
\end{figure}

In this algorithm, a state vector $|\psi\rangle$ is evolved in time according to an effective, but non-hermitian Hamiltonian, $H_\mathrm{eff}$. $H_\mathrm{eff}$ comprises the Hamiltonian $H_{\text{imp}}$ as well as the particle-number conserving terms from the part describing the L/R baths, i.e. the terms proportional to $E^{(\alpha)}_{ij}$ as well as the terms containing the anticommutators in Eq.~\eqref{eq:BathLindbladian}. This deterministic time evolution is interrupted by stochastic jump processes to different particle sectors, mediated by jump operators $L_{k}^{(\beta)}$, see appendix \ref{app:SWF} for details. Observables are determined as the average over expectation values in independent realizations of $|\psi\rangle$.
Such a stochastic unraveling of the Lindblad equation into a pure state description, as described above, only works for proper density operators $\rho$.
When evaluating a Green's function, one needs the stochastic time evolutions operators obtained by applying an operator $A$ to $\rho$.
In order to compute two-time correlation functions, 
\begin{equation}
 G_{BA}(t,t') = \langle \psi(t_0)|B(t)A(t') |\psi(t_0)\rangle
 \label{eq:defGFAB}
\end{equation}
we follow the approach outlined in \tcite{br.ka.97} and consider the stochastic 
time evolution of a doubled Hilbert space resulting in the algorithm in Fig.~\ref{fig: SWFdoubled}.
 
\begin{figure}
\begin{tcolorbox}[title=SWF algorithm in the doubled Hilbert space, halign title=flush center, halign lower=center, fonttitle=\bfseries, lower separated=false, colframe=red!70!white, colback=red!5!white, left=0mm,  leftlower=4mm, sharp corners]
 \begin{enumerate}
  \item Propagate the state $\ket{\psi(t_0)}$ with the SWF algorithm up to a desired time $t'$ and normalize it.
  \item Compute the vector $\ket{\phi(t')} = A\ket{\psi(t')}$ and construct a doubled Hilbert space: \label{it:doubledspace}
  \begin{align*}
    \quad\quad\quad \mathbf{\Theta}(t') = \begin{pmatrix}  \psi(t') \\ \phi(t') \end{pmatrix} &
    \begin{split}
      \mathbf{H}_\mathrm{eff} &= \begin{pmatrix}  H_\mathrm{eff} & 0 \\ 0 & H_\mathrm{eff} \end{pmatrix} \\
      \mathbf{L}_{k}^{(\beta)} &= \begin{pmatrix} L_{k}^{(\beta)} & 0 \\ 0 & \pm L_{k}^{(\beta)} \end{pmatrix}
    \end{split}
  \end{align*}
  \item Record the norm $\mathinner{\lvert\lvert{\mathbf{\Theta}(t')}\rvert\rvert}$, normalize the vector and perform the SWF algorithm with the doubled vector $\mathbf{\Theta}$ and 
operators  $\mathbf{H}_\mathrm{eff}, \mathbf{L}_{k}^{(\beta)}$.
  \end{enumerate}
  \tcblower
  \tcbsubtitle[after skip=\baselineskip]{\centering Green's functions}
  Measure at desired times and average over a sufficient number of realizations $\mathbf{\Theta}_i(t)$.
  \begin{equation}
\label{gba}
    G_{BA}(t,t') = \frac{1}{n} \sum_i^n \frac{\nr2{\mathbf{\Theta}(t')}}{\nr2{\mathbf{\Theta}_i(t)}} \bra{\psi_i(t)} B \ket{\phi_i(t)}
  \end{equation}
\end{tcolorbox}
\caption{The stochastic wave function algorithm in the doubled Hilbert space which allows to calculate correlation functions.}
\label{fig: SWFdoubled}
\end{figure}

Here, a state vector $|\psi\rangle$ is evolved in time together with a corresponding 
vector $A|\psi\rangle$. A Green's function is then proportional to the stochastic sample of off diagonal matrix elements of the second operator $B$, see Eq.~\ref{gba}.
Notice that for single fermion Green's functions, $A$ is a fermionic creation/annihilation operator. In that case one has to use the negative sign in front of the jump term for the lower part of the doubled Hilbert space, cf. Eq.~\ref{eq:L_D_diag}, see appendix B in ~\cite{sc.go.16}. Notice that generalizing the doubled Hilbert space to a multiple Hilbert space allows to sample different correlation functions at once, see Appendix Sec.~\ref{app:Multistates}.\\
The SWF algorithm requires a routine which is able to time-evolve an initial vector with a non-hermitian generator for some (arbitrarily-) small time $dt$\footnote{Although it is in favor of the algorithm if the routine is able to time evolve directly between consecutive jumps, that is for times $dt\approx\tau_{\text{jump}}<1/J$}. In the present work we use the so-called Arnoldi algorithm\cite{kn.ar.11.ec} for the time evolution which is the Lanczos method generalized to the non-hermitian case. For more details we refer to the Appendix Sec.~\ref{app:SWF}.

\section{Extrapolation of Observables to the limit of vanishing cost function}\label{sec:ExtrapolationofObservables}
As illustrated above, AMEA is a method which can be systematically improved by increasing the number of bath sites $N_B$ leading to an exponential decrease in the cost function, $\chi = \sum_{\alpha} \chi_{\alpha}$, which is a measure of the overall accuracy. Clearly, the best approximation for some quantity of interest for given $N_B$ 
is obtained within the auxiliary system with the smallest $\chi$.
To improve on these results one can think of
numerically extrapolating the results to the $\chi\to 0$ limit.
This is equivalent to a scaling to the limit of an infinite number auxiliary bath sites $N_B\to\infty$. However, since the accuracy is directly related to $\chi$ rather than $N_B$, it is more convenient to use $\chi$ as an extrapolating parameter.
For a given observable $A$ of interest we can assume for its deviation from its exact (physical) value 
\begin{equation}
 \Delta A(\chi)  = A_{\text{ph}} - A_{\text{aux}}(\chi)=k_{A}\;\;\chi + O(\chi^2)
 \label{eq:extrapolationA}
\end{equation}
with some constant of proportionality $k_{A}$. This suggests that given a series of value pairs $\{\chi_i,A(\chi_i)\}$ one can obtain an approximation to $A_{\text{aux}}(\chi=0)$ by performing a linear fit in the $(\chi,A)$ plane. Within AMEA a series of value pairs $\{\chi(N_B),A\}$ is naturally generated by the different possible auxiliary system sizes.\\
We want to emphasize that the extrapolation scheme presented here is not able to give a consistent error estimate of the extrapolated value as the uncertainty of the individual data points is unknown and not statistically distributes~\footnote{Here, one has to distinguish between a purely statistical error stemming from the solution of the Lindblad system within stochastic wave functions, which is known and negligible, and the systematic error introduced by the mapping to the auxiliary system which is unknown. Further, the role of the higher order terms in Eq.~\ref{eq:extrapolationA} introduces another source of unknown error. To get a grip on the error due to the AMEA mapping, one could perform the extrapolation in some limit where the true value in the physical system is known, for example at zero interaction strength or for some other parameters where the value is known from the literature. One could then use the deviation from the extrapolation fit as approximation to the error of a data point. Since there is a lot of freedom in obtaining this error estimates - and it will thus be very situation dependent-  we will not pursue this further in the current work where we are interested in an unbiased benchmark of the extrapolation scheme.}. 
Nevertheless, this scheme provides a significant improvement, for example in the current, as can be seen in Fig.~\ref{fig:Current}.

\section{Application to the Interacting Resonant Level Model}\label{sec:ApplicationtotheIRLM}

\begin{figure}
 \includegraphics[width=0.9\columnwidth]{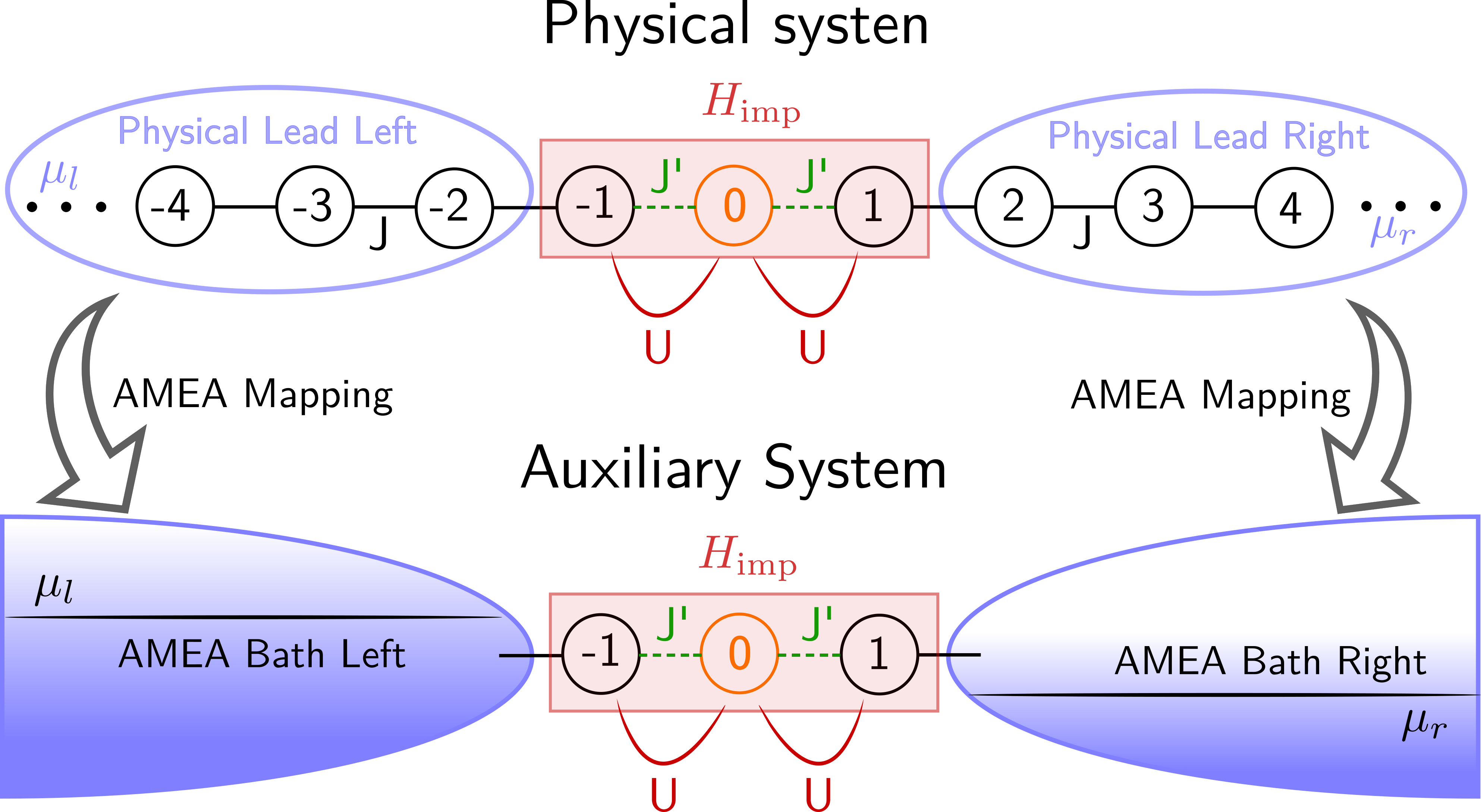}
 \caption{A sketch of the IRLM as lattice model and its mapping to the auxiliary open quantum system used within AMEA.}
 \label{fig:IRLMsketch}
\end{figure}

The IRLM\cite{wi.fi.78} is a commonly used non-equilibrium impurity model of spinless fermions. It features an impurity site connected to two semi-infinite tight-binding chains together with an interaction term coupling the particle densities of the impurity site to the neighboring chain sites, see Fig.~\ref{fig:IRLMsketch}. The Hamiltonian is defined as
\begin{align}
 \notag H_{\text{IRLM}} =& H_{\text{L}} + H_{\text{R}} + H_{\text{dot}}\, , \\
 \notag H_{\text{L}} =& -J\sum_{r = -\infty}^{-2} c_{r}^{\dagger}c_{r+1} + h.c.\, ,\\
 \notag H_{\text{R}} =& -J\sum_{r = 1}^{+\infty} c_{r}^{\dagger}c_{r+1} + h.c. \, ,\\
 \notag H_{\text{dot}} =& -J' \sum_{r=\pm1}c_{r}^{\dagger}c_{0} + h.c.\, ,\\
 &+U\sum_{r=\pm 1} \left( c_{r}^{\dagger}c_{r} - \frac{1}{2}\right)\left(c_{0}^{\dagger}c_{0} - \frac{1}{2}\right)\, ,
 \label{eq:Hamiltonian}
\end{align}
where $c_{r}^{\dagger}/c_{r}$ denote the fermionic creation/annihilation operators at site $r$. Here, $H_{\text{L/R}}$ describe the semi-infinite tight-binding chains of bandwidth $W=4J$ and $H_{\text{dot}}$ introduces the hopping to the impurity as well as the interaction term. A non-equilibrium steady state situation is induced in the system via an applied bias voltage $V$ simulated by shifting the chemical potentials of the leads symmetrically, that is $\mu_l=-\mu_r=\frac{V}{2}$. We use $J$ as unit of energy and work in units where $\hbar=e=k_B=1$.\\
The IRLM is known to be integrable~\cite{wi.fi.78} and becomes equivalent to the continuum model in the so-called scaling regime where the bandwidth becomes the dominant energy scale in the system. Most notably, there is a closed form expression for the steady state current as a function of the bias voltage~\cite{bo.sa.08,ca.ba.11} for the special value of the interaction $U=2$,
\begin{equation}
I(V) = \frac{V}{2\pi} _{2}F_{3}\left[\left\{ \frac{1}{4},\frac{3}{4},1 \right\}, \left\{ \frac{5}{6},\frac{7}{6} \right\}; -\left(\frac{V}{V_c}\right)^6 \right]\, ,
\label{eq:Currentexpression}
\end{equation}
with $V_c=r(J')^{\frac{4}{3}}$ and $r\approx 3.2$\footnote{In more detail, $V_c = \frac{\sqrt{3}}{4^{2/3}} \frac{4\sqrt{\pi}\Gamma (2/3)}{\Gamma (1/6)} T_B$ and $T_B = c(J')^{\frac{4}{3}}$ with $c\approx 2.7$ from \cite{bo.sa.08}.}. Here, $_{2}F_{3}(a,b;z)$ is the generalized hypergeometric function. The formula Eq.~\ref{eq:Currentexpression} is valid at zero temperature and in the scaling regime, where $V,J',U\ll W$\footnote{From previous works~\cite{bi.mi.17}, we know that at $U=2$ one has to restrict to $J'\lesssim 0.5$ and $V\lesssim 2$ to be in the scaling regime.}. In this way,  $I/V_c$ becomes a universal function of the scaled voltage $V/V_c$ alone and in particular does not depend on the hybridization strength $J'$.

\subsection{AMEA for the IRLM}\label{subsubsec:AMEAfortheIRLM}
\begin{figure}
 \includegraphics[width=0.9\columnwidth]{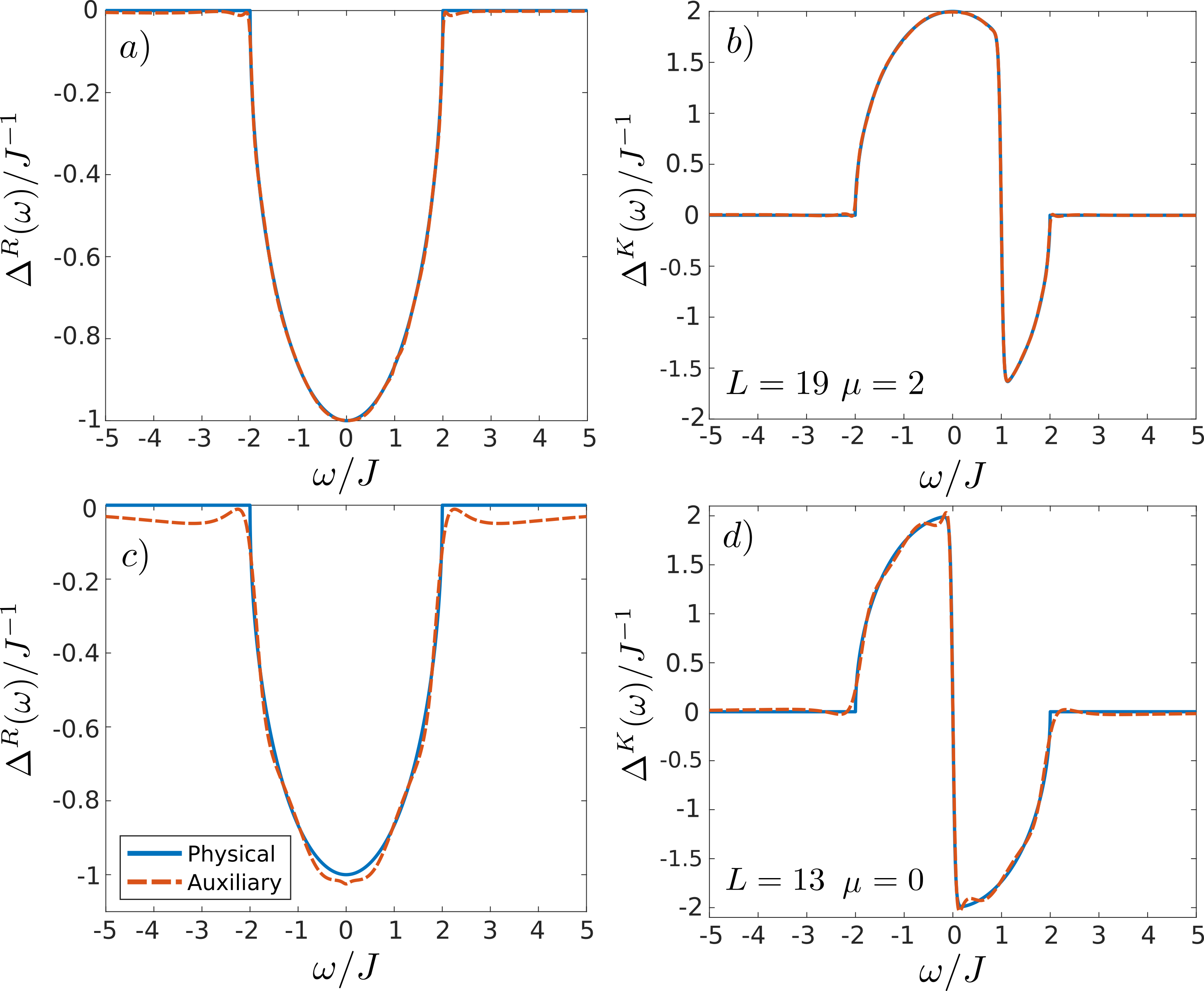}
 \caption{Comparison of the physical and auxiliary hybridization function at the boundary of the  left bath, i.e.  $r=-1$, and $T=0.025$\\
 $a)/b)$ Retarded/Keldysh part of the hybridization function
 for $L=19, \mu=2$, $c)/d)$ Retarded/Keldysh part of the hybridization for $L=13, \mu=0$. 
The $L=19$ results where obtained with the ADAM routine from Sec.~\ref{subsubsec:fit} while $L=13$ was optimized with PT. Solid lines represent the hybridization of the physical system, $\Delta_\text{ph}$, and dashed lines that of the auxiliry system,  $\Delta_\text{aux}$. Panel $a)/b)$ show a fit for $\mu\neq0$ to examplify the capability of representing a non-equilibrium situation. Panel $c)/d)$ illustrate the fit used for the calculation of the equilibrium spectral functions in Fig.~\ref{fig:DOSequilibrium}.}
 \label{fig:Fits}
\end{figure}

In the IRLM, the interaction lives on the contact links to the leads and, therefore,  the interacting region comprises the sites $r=\{-1,0,1\}$ which corresponds to having
\begin{align}
 H_{\text{imp}}&=H_{\text{dot}}\\
 H_{\text{B}_{L}}&=-J\sum_{r = -\infty}^{-3} c_{r}^{\dagger}c_{r+1} + h.c.\\
 H_{\text{B}_{R}}&=-J\sum_{r = 2}^{\infty} c_{r}^{\dagger}c_{r+1} + h.c.\\
 H_{\text{Hyb}}&= -J\left( c_{-2}^{\dagger}c_{-1} + c_{1}^{\dagger}c_{2}\right) + h.c.
 \label{eq:AMEAHmaforILRM}
\end{align}
as indicated in Fig.~\ref{fig:IRLMsketch}. Since $H_{\text{B}_{L/R}}$ describe semi-infinite tight-binding chains in equilibrium, $\Delta_{\text{ph,L/R}}$ represent baths with a semicircular density of states with a bandwidth of $W=4$ and an electronic distribution function given by the Fermi-function. Within AMEA, a given parameter set $E^{\alpha},\Gamma^{\alpha,(1,2)}$ fixes both the density of states as well as the distribution function of the corresponding bath.
Since the Hamiltonian Eq.~\ref{eq:Hamiltonian} is  particle-hole symmetric, it suffices to perform the fit only for one of the two baths, e.g. the left ones, and obtain the parameters of the right bath by particle-hole transformation. Thus, also the cost function for the left and right bath will be equal for a given bias voltage, $\chi_{L}=\chi_R$. To illustrate the mapping, we show in Fig.~\ref{fig:Fits} two examples for such a fit with $L=13$ ($N_B=6$) and $L=19$ ($N_B=9$). 
Notis that the same fit can be used for any set of parameters in this model.

\subsection{Extrapolation of the steady state Current}\label{subsec:SteadystateCurrent}
\begin{figure*}
\includegraphics[width=0.9\textwidth]{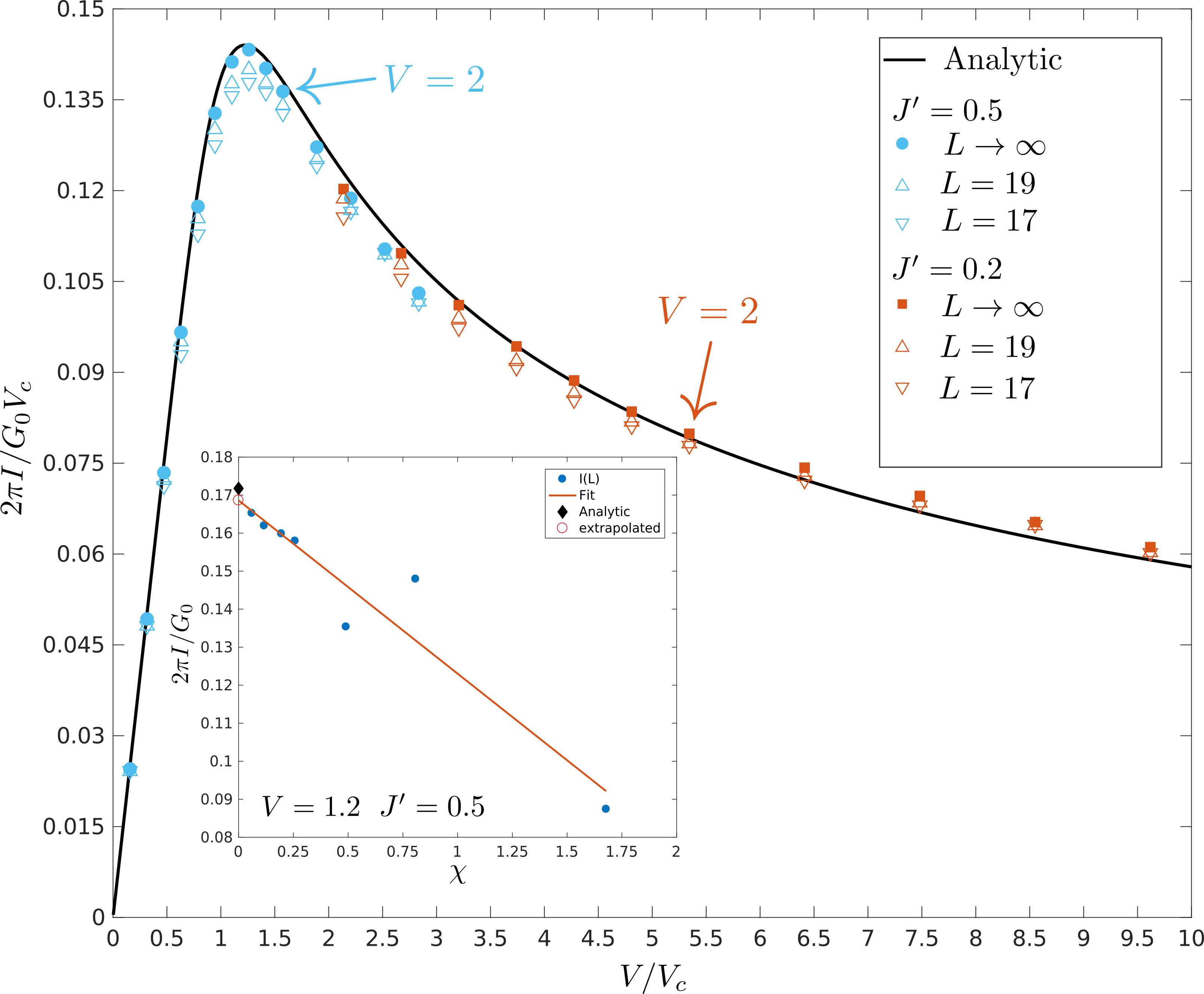}
\caption{Scaled steady state current as function of the scaled bias voltage $V/V_c$. We plot the analytic solution for $T=0$ (solid black line), the extrapolated AMEA current (filled circles), and the current for $L=17$ and $L=19$ (open symbols). Shown are results for $J'=0.2$ (red symbols) and $J'=0.5$ (blue symbols).
The arrows indicate the data points which correspond to the voltage $V=2$ for the two different considered $J'$. The inset shows an example  of the Current vs. cost-function
 $I(\chi)$ for $V=1.2$, $J'=0.5$ (filled blue circles) and the corresponding linear fit (solid red line) as well as the extrapolated value at zero cost-function (open red circle) together with the analytic result (filled black diamond). Other parameters are $T=0.025$ and $U=2$. $G_0=e^2/h$ is the conductance quantum for spinless fermions.}
\label{fig:Current}
\end{figure*}

Since there are no free parameters in Eq.\ref{eq:Currentexpression} we can use this as a benchmark for our numerical approach and test the extrapolation scheme of Sec.~\ref{sec:ExtrapolationofObservables}. However, it should be noted that our results are obtained for $T=0.025$ while Eq.~\ref{eq:Currentexpression} is the result for zero temperature.\\
Given an auxiliary system of size $L$ we can evaluate the  current over a physical bond $i$  in the auxiliary system\footnote{In practice, we measure at all physical bonds and average accordingly. There is a tiny breaking of current conservation due to the numerics.}
\begin{align}
\notag I_{i,i+1} &= E_{i+1,i}\langle c_{i+1}^{\dagger}c_{i}\rangle - E_{i,i+1}\langle c_{i}^{\dagger}c_{i+1}\rangle \, ,\\
 &= 2E_{i+1,i}\iim\langle c_{i+1}^{\dagger}c_{i}\rangle \, ,
 \label{eq:currentfromexpectationvalue}
\end{align}
where the parameters $E_{i,i+s}$ represent the hopping along the chain in the interacting region.
In the following, we consider 
results obtained with
 $7\leq L \leq  19$. In Fig.~\ref{fig:Current} we plot the universal steady state current together with the corresponding data points obtained with AMEA for $J'=0.5$ and $J'=0.2$. Shown are the AMEA results for individual system sizes as well as the extrapolated current.
 For $J'=0.2$ the hybridization strength $\Delta(\omega=0) = J'^2 = 0.04$ becomes comparable to the temperature used in our calculations,  $T=0.025$. Thus, for $J'=0.2$ we disregard data points corresponding to small voltages $V\leq 0.6$ because they are significantly altered by the finite temperature present in the auxiliary system.\footnote{At small voltages, the current is carried by states around the chemical potentials which are most affected by the finite temperature.}\\
We see that the current improves significantly towards the analytic solution thanks to the extrapolation scheme. As discussed above, the analytic solution is only valid for not too large bias voltages\cite{bi.mi.17}. Indeed, we see a systematically growing deviation between the analytic solution and the current from AMEA\footnote{Of course, in the present case it is not strictly possible to distinguish between deviations coming from the finite temperature and ones originating from leaving the scaling regime.} for 
voltages $V\gtrsim 2$, see the markers in Fig.~\ref{fig:Current}\\
The inset in Fig.~\ref{fig:Current} shows an example extrapolation. As one would expect, the data points with bigger cost functions (smaller system sizes) show a stronger scattering from the linear fit than the more accurate points. While the points with low cost functions make for more confidence in the results, the accuracy of the extrapolated current does not suffer when the biggest system size, $L=19$, is excluded from the analysis.
This suggests that when utilizing the extrapolation to zero cost function, it is probably not necessary to simulate the biggest system sizes within reach. Rather, one can check for a small fraction of points whether or not the - usually very cpu-time intensive - bigger system size(s) are worth calculating\footnote{If error estimates are used, points at lower cost-functions will reduce the uncertainty in the final result.}.

\subsection{Spectral function of the IRLM}\label{sec:Spectralfunctions}
In this section, we evaluate 
 the steady-state single-particle Green's function $G$  at the central impurity site.
The calculation is carried out in the real time domain and we use the approach discussed 
in Sec.~\ref{sec:SWF}, see also Sec.~\ref{app:Practicalimplementation}. We use a step size of $dt=0.05$ and $10^5$ time steps to first reach the steady state at $t_0=5*10^3$. We have verified that expectation values of static observables don't change after this time. Then we sample the Green's function $G(t - t_{0})$ for later times beyond $t_0$ up to $t_{\text{end}} = t_0 + 6000dt$. This is sufficient, since here $G(t_{\text{end}}-t_{0})<10^{-6}$

\begin{figure}
 \includegraphics[width=0.9\columnwidth]{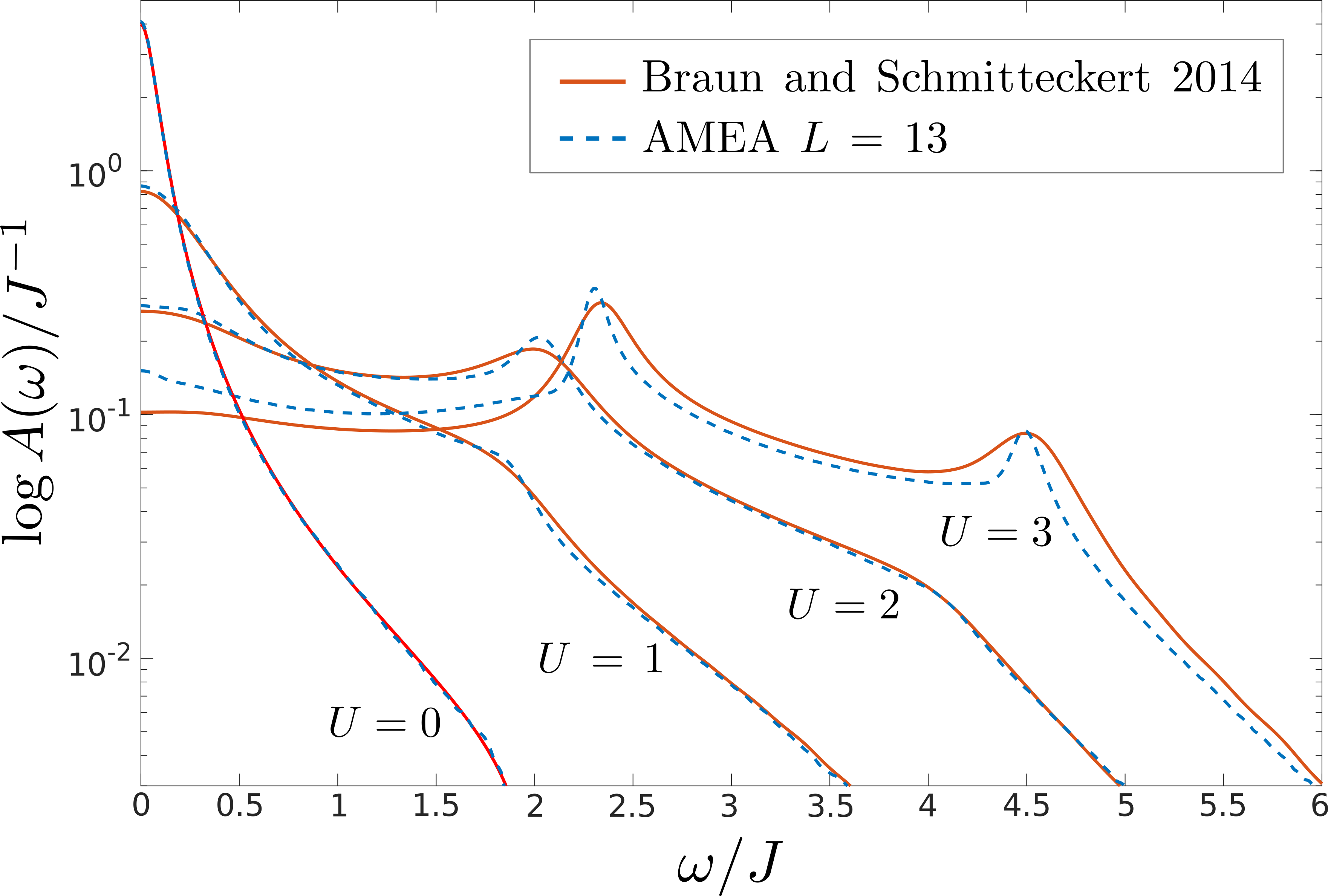}
 \caption{Equilibrium  ($V=0$) spectral function at the impurity site, $r=0$, for different interaction strengths. We compare our results with Braun \etal~\cite{br.sc.14} (obtained at $T=0$). Our parameters are $J'=0.2$, $T=0.025$.}
  \label{fig:DOSequilibrium}
\end{figure}

 Finally, we average $G$ over $O(10^5)$ realizations and determine the spectral function by direct Fourier transform~\footnote{We have tried to take the statistical error into account for the Fourier transform within the framework of Bayesian probability theory but it did not lead to more satisfactory results. It turns out that an accurate  error estimate for a given frequency is simply given by linear error propagation in the numerical Fourier integration}. 
All results presented in this section are obtained with an auxiliary system of size $L=13$. The corresponding hybridization function is shown in the lower panels of Fig.~\ref{fig:Fits}.\\
Like any non-equilibrium approach, AMEA is also applicable in equilibrium situations which is just the special case when $\mu_l = \mu_r = 0$ allowing us to compare our results against the literature. In Fig.~\ref{fig:DOSequilibrium}, we compare our results to the equilibrium density of states obtained by Braun and Schmitteckert via MPS~\cite{br.sc.14}.
For interaction strengths $U<2$ that are small compared to the bandwidth, we observe a very good agreement with the reference over the whole frequency range. At the self dual point $U=2$ we start to see small quantitative deviations of peak heights but still obtain an satisfactory agreement. When the interaction becomes comparable to the bandwidth, $U=3$, the deviations become significant and continue to grow as the interaction is increased (not shown). The reason for the growing deviations is that in the present AMEA mapping the region outside the bandwidth is not well reproduced, see also Fig.~\ref{fig:Fits}. While these states do not play a role as long as all energy scales in the system are small compared to the bandwidth, i.e. in the scaling regime, the details of the leads at higher energies become important when the interaction becomes comparable to the bandwidth. The latter does, however, not mean, that AMEA is not at all applicable in this parameter regime, rather one has to make sure that the region outside the bandwidth is also faithfully reproduced by the auxiliary system. This can be achieved by using a differently distributed cost function in the fit or by going to larger auxiliary system sizes.\\

\subsection{Performance:}
From a numerical point of view, the stochastic wave function (SWF) method has two main advantages. First, since one evolves wave functions there is no need to square the Hilbert space as when one deals with the density matrix. This means that one can use a twice as large $L$, and thus, achieve a much better accuracy.~\footnote{This does not hold for approaches in which the system size is not a limitation, such as tensor network states, i.e. MPS, where the entanglement entropy encoded into the state limits the simulation.}. 
Second, individual realizations of possible time evolutions are independent which means that the method is easily parallelizable. This makes SWF very suitable for future cluster facilities which thrive on highly parallel algorithms.

However, the prize to pay is a cpu time that is about twenty times longer than solving an auxiliary system with the same value of the cost-function by MPS\footnote{We note that the size $L$ of the auxiliary system solved by MPS must be larger, compared to the present approach, to reach the same value of the cost-function. This is because in MPS, the matrices $E,\Gamma^{(1)},\Gamma^{(2)}$ have to be tridiagonally restricted which lowers the number of fit parameters available for a given $L$.}. On the other hand, thanks to parallelization, the  wall-time\footnote{The time it takes before the result is known} can obviously 
be made almost arbitrarily small. For example, the GF's for $L=13$  in Fig.\ref{fig:DOSequilibrium} where averaged over about half a million realizations where a single one takes around one second. For comparison, the solution with the super-fermion plus ED approach for $L=13$ would be in the order of minutes.

\section{Summary, Conclusion and Outlook}\label{sec:Conclusion}
We reported on technical developments within the auxiliary master equation approach and applied it to the 
Interacting Resonant Level Model (IRLM) in and out of equilibrium to benchmark the new techniques. We successfully applied the Stochastic Wave Function (SWF) algorithm to determine the steady state properties of the auxiliary Linblad system. On the one hand the SWF algorithm is highly parallelizable allowing to reach very low wall-times. On the other hand, we found that in the current implementation of SWF+ED the total cpu-time for a spectral function is twenty times higher than in available alternatives for the solution of the auxiliary system introduced by AMEA. Further, we saw that an auxiliary system size of $L=13$ is enough to obtain reliable spectral information of the IRLM for interactions $U\lesssim W/2$.

We obtained a further significant improvement by extrapolating physical quantities, most notably the current, to the $N_B\to \infty$ limit. In fact, it turns out to be more effective to extrapolate linearly in the cost function $\chi$, which then would correspond to an exponential extrapolation in $N_B$. Such an extrapolation is able to improve the results significantly and possibly circumvents the need to go to larger system sizes.

In addition, we introduced a variable rank parametrization of the auxiliary Lindblad matrices which typically reduces the number of fitting parameters in the AMEA mapping. Employing the new parametrization together with an optimization routine from machine learning, we were able to maintain an exponential decrease of the cost function also for larger system sizes where the previously used parallel tempering algorithm failed

In the current work, we calculated spectral functions only in equilibrium, where we can compare to the literature, while the present method also allows to calculate spectral functions in the non-equilibrium situation. Since the current through a system can also be expressed in terms of GF's, we can investigate how the negative differential conductance in the IRLM arises from the spectral properties out of equilibrium. However, this is beyond the scope of the present work and will be presented elsewhere.

Further improvement in accuracy and computational time could be possibly be achieved by linear prediction, in order to extrapolate the Green's function to large times, and by combining MPS with the present SWF approach.

\begin{acknowledgments}

We would like to thank Irakli Titdvinidze, Daniel Bauernfeind and Gerhard Dorn for fruitful discussions.  A special mention goes to Franz Scherr who introduced us to the machine learning environment tensorflow and provided a first implementation for the AMEA mapping. We are grateful to Peter Schmidteckert for providing us with the reference data for the spectral functions.
This work was partially supported by the Austrian Science Fund (FWF)  within 
Projects  P26508 and F41 (SFB ViCoM),  as well as NaWi Graz.
The calculations were partly performed on the dCluster and lCluster Graz 
as well as the VSC-3 cluster Vienna

\end{acknowledgments}

\appendix

\section{Technical details of the SWF algorithm}\label{app:SWF}
In order to present the SWF algorithm, we consider a general Lindblad system for a generalized ``density-matrix'' $\tilde{\varrho}=f(\{c^{(\dagger)}\})\rho$ where $f(\{c^{(\dagger)}\})$ denotes some function of fermionic operators,

\begin{equation}
 \li = \li_{H} + \li_{D} \, .
 \label{eq:Ltot}
\end{equation}
It is composed of a central region with Hamiltonian $H$ and the corresponding Liouvillian $\li_{H}$, 
\begin{equation}
 \li_{H} \tilde{\varrho}= -i[H,\tilde{\varrho}] \, ,
 \label{eq:L_H}
\end{equation}
and a dissipative part described by $\li_{D}$,
\begin{equation}
\begin{split}
 \li_D \tilde{\varrho} &= 2 \sum_{ij} \ga1_{ij} \left( \pm c^\nag_{j}\tilde{\varrho} c_{i}^\dagger - \frac{1}{2} \left\{\tilde{\varrho}, c_{i}^\dagger c^\nag_{j} \right\} \right)  \\ 
 &+ 2 \sum_{ij} \ga2_{ij} \left( \pm c_{i}^\dagger\tilde{\varrho} c^\nag_{j}  - \frac{1}{2} \left\{\tilde{\varrho}, c^\nag_{j}c_{i}^\dagger \right\} \right)\, .
\end{split}
\label{eq:L_D}
\end{equation}

Here, $i$ and $j$ run over all $L$ sites of the system and $\Gamma^{(1)/(2)}$ are $L \times L$ matrices.
The minus sign in $\Eq{eq:L_D}$ is valid, if $\tilde{\varrho}$ is odd in the number of fermion operators, i.e $\tilde{\varrho} = c^{(\dagger)}_i\rho$.
This is the case with Green's functions, where we need to propagate $c_i^{(\dagger)}\rho$.

In order to obtain the jump operators one has to
 diagonalize the matrices $\ga\beta$, $\beta = 1,2$,

\begin{equation}
 2\ga\beta_{ij} = \sum_k U^{(\beta)}_{ik}\gamma_k^{(\beta)}U^{(\beta)^*}_{jk} \nonumber \, ,
\end{equation}

and end up with the eigen-decomposition of the dissipator,

\begin{align}
 \li_D \rho &= \sum_{\beta k} \left( \pm L_{k}^{(\beta)} \rho L^{(\beta)\dagger}_{k} - \frac{1}{2}\left\{\rho, L^{(\beta)\dagger}_{k}L^{(\beta)}_{k}\right\} \right) \label{eq:L_D_diag} \\
\begin{split}
 L^{(1)}_{k} &= \sum\limits_i \sqrt{\gamma_k^{(1)}}U^{(1) *}_{ik}c_{i}  \\
 L^{(2)}_{k} &= \sum\limits_i \sqrt{\gamma_k^{(2)}}U^{(2)}_{ik}c_{i}^\dagger\, .
 \label{eq:jumpoperators}
\end{split}
\end{align}

The anti-commutators in \eq{eq:L_D_diag}  are included into the effective, non-hermitian Hamiltonian
\footnote{For this generalize $[H,\rho]$ to $H_\mathrm{eff}\rho - \rho H_\mathrm{eff}^\dagger$.}

\begin{equation}
\label{heff}
 H_\mathrm{eff} = H - \frac{i}{2} \sum_{\beta k}  L^{(\beta)\dagger}_{k}L^{(\beta)}_{k}
\end{equation}

With this Hamiltonian and the jump operators $L_{k}^{(\beta)}$, \Eq{eq:jumpoperators}, one formulates the SWF algorithms in Sec.~\ref{sec:SWF}, Fig.~\ref{fig: SWFalgorithm} and Fig.~\ref{fig: SWFdoubled}.

\subsection{Jump-time search and Arnoldi}\label{app:jumptimesearch}
As mentioned in Sec.~\ref{sec:SWF} we use the so-called Arnoldi algorithm~\cite{kn.ar.11.ec} for the time evolution. Arnoldi  is a Krylov space method analogue to Lanczos but for non-hermitian Hamiltonians. For a given initial state, $|\psi_0\rangle$ and time interval $dt$, a Krylov space, spanned by $Q$, is generated by iteratively applying $H_\mathrm{eff}$ to the starting vector until a satisfactory approximation for the time evolution operator $e^{-iH_\mathrm{eff}dt}\approx Q^{\dagger}e^{-iH_\mathrm{K}dt}Q$ is found. For any given time $t$ up to the maximal time $dt$, the state and the corresponding norm needed for the SWF algorithm are given by
\begin{align}
  |\psi(t)\rangle &= Q^{\dagger} e^{-iH_\mathrm{K}t}Q|\psi_0\rangle\ = Q ^{\dagger}e^{-iH_\mathrm{K}t}\vec{v}_0\, , \\ 
  \vec{v}_0 &= Q|\psi_0\rangle = (1,0,0,...)^{\top}\, ,\\ 
 \|\psi(t)\|^2 &=\langle\psi_0|Qe^{iH_\mathrm{K}^{\dagger}t}\underbrace{Q Q^{\dagger}}_{\mathds{1}}e^{-iH_\mathrm{K}t}Q|\psi_0\rangle\, , \\ 
  &=\vec{v}_0^{\top} e^{iH_\mathrm{K}^{\dagger}t}e^{-iH_\mathrm{K}t}\vec{v}_0\, ,
 \label{eq:arnolditime-evo}
\end{align}
where we have used the property that $Q|\psi_0\rangle$ is nothing else than the first Krylov vector and $Q Q^{\dagger}=\mathds{1}$ is the identity\footnote{Note that for non-hermitian problems $Q^{\dagger}Q\neq\mathds{1}$}. We want to point out that by virtue of Eq.~\ref{eq:arnolditime-evo} the norm can be calculated within the Krylov space representation itself, which is typically of size $\text{dim}_{K}=O(10)$, without the need to use the transformation matrices $Q$ which are of dimension $\text{dim}_Q = \text{dim}_F \cdot \text{dim}_K$ where $\text{dim}_F$ is the dimension of the Hilbert space (many-body Fock space). Differentiating Eq.~\ref{eq:arnolditime-evo}, yields
\begin{equation}
 \frac{d}{dt}\|\psi(t)\|^2 = -2i\iim\left(\vec{v}_0^{\top}e^{iH_\mathrm{K}^{\dagger}t}H_\mathrm{K}e^{-iH_\mathrm{K}t}\vec{v}_0\right)\, ,
 \label{eq:diffnormt}
\end{equation}
which allows to determine the jump time $t_{j}$ in the SWF algorithm, satisfying $\|\psi(t_{j})\|^2 - r_{j} = 0$, by applying Newtons method.

\subsection{Practical implementation for the steady state situation}\label{app:Practicalimplementation}
Here we want comment on the practical implementation for the special case of steady state quantities.
\subsubsection{Steady state observables}
We start with the simpler case of sampling a steady state observable. A steady state expectation value is obtained like in a Monte Carlo (MC) simulation. We start with a random starting state and time evolve the system until it reaches the steady state, where the system is time-translational invariant (like the thermalization in a MC simulation). Once we are in the steady state, we start measuring the observable generating an autocorrelated time series from which an estimator of the expectation value can be obtained. As usual the time-series needs to be long enough to have overcome autocorrelations, which can be checked for example by a Binning plot.\\
For the present case we typically recorded $N_m = 2^{18}$ measurements separated by a time $\Delta t = N_{t_\text{skip}} dt$ with a time-step $dt = 0.05$ and $N_{t_\text{skip}} dt = 16 dt \approx 10 \bar{t}_j$, where  $\bar{t}_j$ is the average jump time. For thermalization we performed additionally $10\%$ of the total time evolution leading to $O(10^5)$ thermalization time-steps.
Parallelization can be achieved by running several individual walkers on a single cluster node, where each walker is bound to one core for instance.

\subsubsection{Steady state single particle GF's}
To obtain steady state GF's of the Lindblad system we follow Ref~\cite{do.ga.15}. In short, it is best to calculate the lesser and greater steady state GF, defined by
\begin{equation}
 G_{ij}^<(t) = i\langle c_i^\dagger(t)c_j\rangle_{\infty}, \hspace{20pt} G_{ij}^>(t) = -i\langle c_i(t)c_j^\dagger\rangle_{\infty}
\label{eq:defGFlessgrtr}
\end{equation}
where $\langle \cdot \rangle_{\infty} = T_r\{\cdot \rho_{\infty}\}$ denotes the expectation value in the steady state. We sample the GF by first time-evolving into the steady state like above. Next, we apply the operator $c_r^{(\dagger)}$, construct the doubled Hilbert space, continue to time-evolve in the doubled Hilbert space and measure according to the SWF algorithm in the doubled Hilbert space.\\
As stated in the main text the time steps needed for GF's is of $O(10^3)$ and to reach the accuracy needed for smooth spectral functions, we had to average over $O(10^5)$ realizations. Further, we perform $O(10^5)$ time-steps to get into the steady state.
For the performance in terms of cpu-time, it is crucial that the time steps into the steady state are done only for a small fraction of the realizations; the corresponding final states are saved\footnote{In the present case the time-steps into the steady state make for about ten per cent of the total run time.}. Another realization starts from a state obtained by time evolving such a saved state for some time $\Delta t\approx 100\bar t_{\text{j}}$, to make sure that individual realizations are independent to a very good approximation\footnote{Only early times will be correlated as the realizations gain in independence through the jumps in the time-evolution. One can test for autocorrelations when considering the different realizations for a specific (early-) time step as a time series and apply autocorrelation analysis.}.
\paragraph{Multistates:}\label{app:Multistates}
One can sample multiple correlation functions, $G_{B_iA_i}(t,t')$, together when generalizing the doubled Hilbert space to a multiple Hilbert space. For this, generalize
\begin{equation}
\Theta(t) = \begin{pmatrix}  \psi(t) \\ \phi_1(t) \\ .\\.\\.\\\phi_n(t)\end{pmatrix}
\end{equation}
with the excited states $\phi_i = A_i\ket{\psi}$. For instance, this allows to sample the lesser and greater GF together in a tripled Hilbert space or multiple components of a cluster GF. The advantage is that $|\psi\rangle$ is only time evolved ones, where as in the individual approach, with only a doubled Hilbert space, $|\psi\rangle$ is time evolved $n$-times.
\newline
\paragraph{Destroyed states in the multiple Hilbert space}\label{app:Destroyedstates}
Here, we want to elaborate on the fact that part of the state may be destroyed when applying the SWF algorithm in the multiple Hilbert space. For simplicity, we consider in the following a doubled Hilbert space. Part of the state can get destroyed, when the system leaves the physical particle sectors through the application of a jump operator\footnote{Naturally, this happens more often in smaller systems.}. For instance, a state can get destroyed when the system is in the $N=L$ particle sector and a jump operator $L^{(2)}_{k}$ gets chosen that increases the particle number.\\
First, let us note that this cannot happen in the single Hilbert space since the corresponding weight $w_{\beta k} \propto \nr2{ L_{k}^{(2)}\psi(t_j)}$ is zero and this jump operator will never be chosen.\\
The situation is different in the doubled Hilbert space when the two components of a state reside in different particle sectors. Too see this, let us consider the case of the greater GF. Here, if $|\psi\rangle$ is in sector $N$, $|\phi\rangle$ will always describe a state with $N+1$ particles, since the jump operator applied is the same for both components. If at some time $t_{\text{kill}}$, $|\phi\rangle$ is in the sector $L$, the weight for a jump operator that increases the particle number, $w_{\beta k} \propto \nr2{ L_{k}^{(2)} \psi(t_j) } + \nr2{ L_{k}^{(2)}\phi(t_j)}$, might be non-zero since the first part can be non-vanishing.\\
If part of the state is destroyed, all subsequent measurements in this specific realization of the time series for the GF will all be zero.\\
It is important to realize that this is the correct behavior. It exemplifies why the doubled Hilbert space is needed when calculating correlation functions and why it would be wrong to simply consider an independent time evolution for the excited state and the initial state separately. In fact, in the independent approach, any correlation between the initial state and the final state would be lost very quickly through the stochastic process and it is key that the two states always jump together, thereby mediating the correlation.

 \bibliographystyle{./prsty}
 \bibliography{Additional_Refs.bib,references_database.bib}
\end{document}